\documentclass[useAMS,usenatbib]{mn2e}
\usepackage{graphicx}
\usepackage{txfonts}
\usepackage{natbib}

\title[\dgcvn]{ Early optical follow-up of the nearby active star {\dgcvn} during its 2\,014 superflare }

\author[Caballero-Garc\'{i}a et~al.]{M.~D. Caballero-Garc\'{i}a$^1$\thanks{E-mail:
cabalma1@fel.cvut.cz},  V. \v{S}imon$^{1,2}$, M. Jel\'{\i}nek$^{3}$, A.~J. Castro-Tirado$^{3,4}$, \newauthor
A.~Cwiek$^{5}$, A. Claret$^{3}$, R.~Opiela$^{6}$, A.~F. \.Zarnecki$^{7}$, J. Gorosabel$^{3,8,9}$, \newauthor
S.~R. Oates$^{3}$, R. Cunniffe$^{3}$, S. Jeong$^{3,14}$, R. Hudec$^{1,2}$, V.~V. Sokolov$^{10}$, \newauthor
D.~I. Makarov$^{10}$, J.~C. Tello$^{3}$, O. Lara-Gil$^{3}$, P. Kub{\'a}nek$^{11}$, S. Guziy$^{12}$, \newauthor
J. Bai$^{13}$, Y. Fan$^{13}$, C. Wang$^{13}$, I.~H. Park$^{14}$  
\\
\\
$^{1}$ Czech Technical University in Prague, Faculty of Electrical Engineering, Technick\'a 2, 166 27  Praha 6, Czech Republic \\
$^{2}$ Astronomical Institute, Academy of Sciences of the Czech Republic, 251~65~Ond\v{r}ejov, Czech Republic \\
$^{3}$ Instituto de Astrof\'{\i}sica de Andaluc\'{\i}a (IAA-CSIC), P.O. Box 03004, E-18080, Granada, Spain \\
$^{4}$ Unidad Asociada Departamento de Ingenier\'{\i}a de Sistemas y Autom\'atica, E.T.S.  de Ingenieros Industriales, Universidad de M\'alaga, Spain \\
$^{5}$ National Centre for Nuclear Research, Ho\.za 69, 00-681 Warsaw, Poland \\
$^{6}$ Center for Theoretical Physics of the Polish Academy of Sciences, Al. Lotnikow 32/46, 02-668 Warsaw, Poland \\
$^{7}$ Faculty of Physics, University of Warsaw, Pasteura 5, 02-093 Warszawa, Poland \\
$^{8}$ Asociada Grupo Ciencia Planetarias UPV/EHU-IAA/CSIC, Departamento de F\'{\i}sica Aplicada I, \\
E.T.S. Ingenier\'{\i}a, Universidad del Pa\'{\i}s Vasco UPV/EHU, Alameda de Urquijo s/n, E-48013 Bilbao, Spain \\
$^{9}$ Ikerbasque, Basque Foundation for Science, Alameda de Urquijo 36-5, E-48008 Bilbao, Spain \\
$^{10}$ Special Astrophysical Observatory of R.A.S., Karachai-Cherkessia, Nizhnij Arkhyz, 369167 Russia \\
$^{11}$ Fyzik\'aln\'i \'ustav AV \u{C}R, v. v. i. Na Slovance 1999/2, 182 21 Praha 8, Czech Republic \\
$^{12}$ Nikolaev National University, Nikolska 24, Nikolaev, 54030, Ukraine \\
$^{13}$ Yunnan Astronomical Observatory, Chinese Academy of Sciences, Kunming 650011, Yunnan, China \\
$^{14}$ Department of Physics, Sungkyunkwan University, Suwon, Korea
}

\def\dgcvn{\hbox{\rm DG~CVn}}

\begin{document}


\pagerange{\pageref{firstpage}--\pageref{lastpage}} \pubyear{2002}

\maketitle

\label{firstpage}

\begin{abstract}
{\dgcvn} is a binary system in which one of the components is an M type dwarf ultra fast
rotator, only three of which are known in the solar neighborhood. 
Observations of {\dgcvn} by the {\it Swift} satellite and several
ground-based observatories during its super-flare event on 2\,014 allowed us to perform a complete hard X-ray -- optical
follow-up of a super-flare from the red-dwarf star. The observations support the fact that the super-flare
can be explained by the presence of (a) large active region(s) 
on the surface of the star. Such activity is similar to the most extreme solar 
flaring events. This points towards a plausible extrapolation between the behaviour from the most active red-dwarf stars
and the processes occurring in the Sun. 
\end{abstract}


\section{Introduction}

On April 23rd 2014, at 21:07:08 UT one of the stars from {\dgcvn} flared bright enough (300
milliCrab in the 15-150\,keV band) to trigger the {\it Swift}
satellite's \citep{gehrels04} Burst Alert Telescope (BAT; \citealt{barthelmy05}).
Within two minutes of this ${T}_{0}$, {\it Swift} had slewed to point its
narrow-field telescopes to the source, which revealed gradually decreasing soft
X-ray emissions with a second weaker flare occurring at ${\rm T}_{0}+11$\,ks,
followed by several smaller flares. This behaviour was observed in both the optical
and X-ray bands. On the ground, the wide-field ``Pi of the Sky'' \citep{cwiok07,mankiewicz14} (PI) instrument was observing, covering {\it
Swift}'s field of view, and recorded the optical behaviour of {\dgcvn} even
before the burst began and continued until ${\rm T}_{0}+1100$\,s.
The {\it BOOTES}--2 \citep{castro99,castro12} telescope began to observe {\dgcvn} from ${\approx}{\rm
T}_{0}+11$\,min, starting to take spectra later at ${\approx}{\rm T}_{0}+1$\,h with the low-resolution spectrograph COLORES \citep{rabaza14}, covering the period of
the second flare. Observations with {\it BOOTES}--2 continued for several weeks following
the trigger. Deeper spectra were obtained later with instruments/spectrographs on larger telescopes: OSIRIS \citep{cepa00}
on the {\it Gran Telescopio de Canarias} (GTC) at ${\rm T}_{0}+1.2$\,d,
SCORPIO \citep{afanasiev05} on the 6\,m BTA-6 telescope at SAO in the Caucasus at ${\rm
T}_{0}+15$\,d and CAFE \citep{aceituno13} on the 2.2\,m telescope at Calar Alto at ${\rm
T}_{0}+53$\,d. Fig.~1 shows the data from the first flare: the optical lightcurve from PI
together with the 15-25\,keV {\it Swift} data, from ${\rm T}_{0}-150$\,s to
${\rm T}_{0}+300$\,s. Optical and X-ray observations of the second flare by
{\it BOOTES}--2 and {\it Swift} XRT are shown in Fig.~2.

\begin{figure}
\includegraphics[bb=0 0 612 792,width=7cm,angle=270,clip]{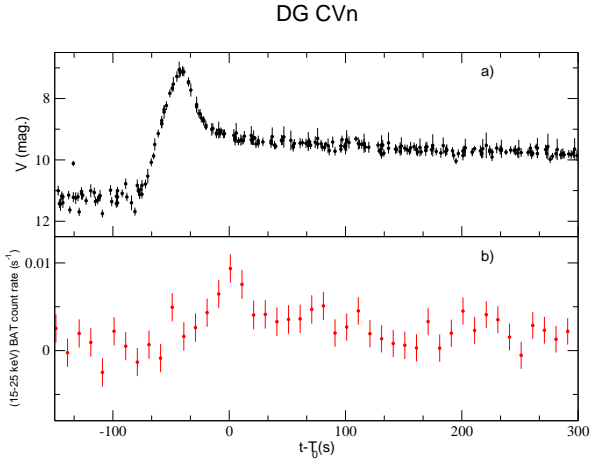}
\caption{{\bf Hard X-ray delay versus the optical emission of the prompt emission (first flare) from {\dgcvn}}. {\bf a}, Optical light
curve from the PI (from ${\rm T}_{0}-150$\,s onwards); {\bf b}, {\it Swift}/BAT light curve in the 15--25\,keV energy range (i.e. hard X-rays). The time (in seconds)
is measured with respect to the BAT trigger time (${\rm T}_{0}$; \citealt{delia14}).
}
\end{figure}

\begin{figure}
\includegraphics[bb=0 0 338 360,width=12cm,angle=0,clip]{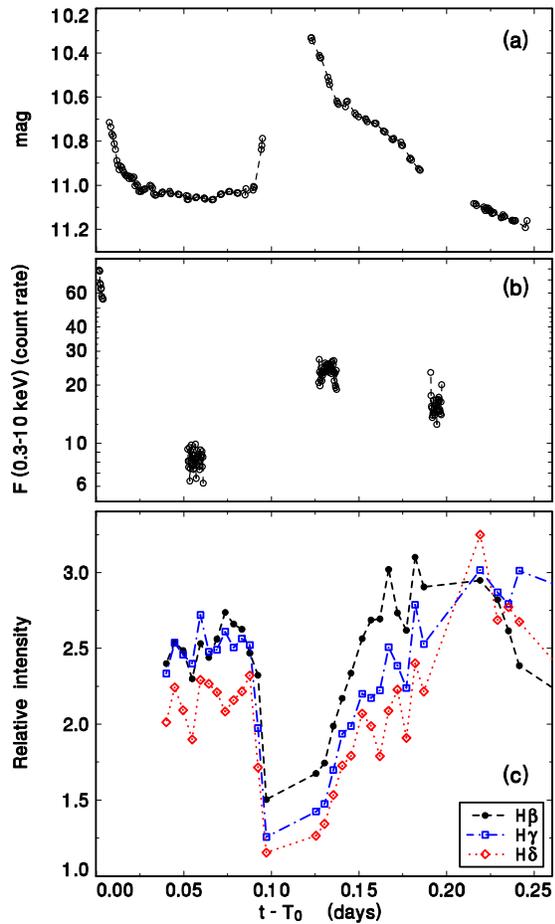}
\caption{{\bf The X-ray brightening was accompanied by an increase of brightness of the optical continuum and the relative intensity of some
spectral lines during the second flare episode}. {\bf a}, Optical {\it BOOTES}--2 light curve in V filter during the second flare
of {\dgcvn} at ${\rm t}-{\rm T}_{0}=0.127$\,d ; {\bf b}, {\it Swift}/XRT light curve in the 0.3--10\,keV energy range (the Flux is measured in ${\rm cts}\,{\rm s}^{-1}$); {\bf c}, Intensities of the Balmer lines (relative to the
continuum) vs. time obtained with the {\it BOOTES}--2/COLORES spectrograph. The uncertainties of the peak intensities are comparable to the size of the symbols. The time
(in seconds) is measured with respect to the BAT trigger time (${\rm T}_{0}$; \citealt{delia14}).
}
\end{figure}

\subsection{\dgcvn}

\noindent DG Canum Venaticorum (DG~CVn; with coordinates (J2000) ${\alpha}=13^{\rm h}31^{\rm m}46.7^{\rm s}$, ${\delta}=29^{\circ}16^{'}36^{''}$; also
named G~165-8AB; \citealt{gliese91} and 1RXS~J133146.9$+$291631; \citealt{zickgraf03}) is a bright (${\rm V}=12.19$; \citealt{xu14}) and
close (${\rm D}=18$\,pc) visual dM4e binary system \citep{riedel14,henry94,delfosse98}. It is also a radio emitting source \citep{helfand99}.
It has been seen that the system has Ca, H and K lines in emission \citep{beers94}. The components
have an angular separation in the sky of $0.17$\,arcsec \citep{beuzit04}, an orbital period of ${\rm P}{\approx}7$\,yr and magnitudes ${\rm V}=12.64,12.93$, for the primary and
the secondary components, respectively \citep{riedel14,weis91}. It is classified as a (joint spectral type from unresolved multiples) M4.0V spectral-type star \citep{riedel14}. It is
also a red high proper motion dwarf-star (${\pi}{\approx}80\,{\rm mas}$; \citealt{jimenez12}).

One of the components of DG~CVn (it is not known which one) has been reported to be chromospherically active \citep{henry94} and one of the only three known M dwarf
ultra-fast rotators in the solar neighborhood. The projected rotational velocity
is $v{\sin(i)}=55.5\,{\rm km}\,{\rm s}^{-1}$, as measured from rotational broadening of the H emission lines seen in high-resolution
spectra \citep{delfosse98}. Recently it has been reported the detection of intense radio emission (the highest ever detected in an active red-dwarf) coinciding
with the time of the first and the second flares reported
in this paper \citep{fender14}. Because the two components of the system are separated by 3.6\,AU (${\approx}2\,500\,{\rm R}_{\star}$) the two stars are not magnetically interacting. Therefore
the intense radio emission has been interpreted as a consequence of the processes occurring in one of the stars (we will be referring to this as {\dgcvn} hereafter). For this
star to rotate so rapidly
a tertiary close companion (i.e. apart from the distant known companion) is expected to exist, but recent studies \citep{fender14} indicate that this might not be the case and that the
youth of the system is the cause. Therefore, this system is considered to be a young star ($30\,{\rm Myr}$; \citealt{riedel14,delfosse98}). Nevertheless it lies outside the spatial
and kinematic boundaries of all known stellar young associations \citep{riedel14}.

\indent {\dgcvn} has been discovered to be an optically variable source \citep{robb99}. The light curve shows periodic sinusoidal variations with a peak to peak amplitude and
photometric period of ${\Delta}\,{\rm R}=0.03$\,mag and ${\rm P}_{\rm phot}=0.10836(2)$\,d, respectively. This variation is thought to be produced by
a hot spot, whose projected area changes as the star rotates. The light-curve was fitted well with a spot of approximately 3.5 degrees
in projected radius with a temperature of 1.3 times the surrounding photosphere. This is the maximum possible value, since the effects of gravity darkening, that might be
important for highly rotating stars, might have not been taken into account.

\section{Observations} \label{observ}

\subsection{X-rays, ${\gamma}$-rays and optical}

\noindent The BAT instrument on-board the {\it Swift} satellite detected a superflare from {\dgcvn} and triggered observations at
2014--04--23, 21:07:08\,UT (i.e. hereafter called ${\rm T}_{0}$; \citealt{delia14}). During this (so-called)
main flare the hard X-ray source had a peak intensity in the BAT
(15-50\,keV) band of ${\approx}300\,$mCrab or $0.06\,{\rm count}\,{\rm cm}^{-2}\,{\rm s}^{-1}$. The initial {\it Swift} X-Ray
Telescope (XRT; \citealt{burrows05}) flux in the 2.5\,s image
was $4.10{\times}10^{-9}\,{\rm erg}\,{\rm cm}^{-2}\,{\rm s}^{-1}$ (0.2-10\,keV). In the optical/UV the source was too bright and saturated {\it Swift}/UVOT during the
first flare \citep{drake14}. An optical counterpart coincident with {\dgcvn} (and also with the X-ray source 2XMM J133146.4+291635) was identified \citep{xu14}
at 21:41:20 UT on 2014--04--24. Initial reports of the optical light-curve indicated significant enhancement in brightness
at the position of the transient and a subsequent fading of 0.92\,mag/hour for the first hour of observation
after the peak (23:30-00:30 UT), followed by a decay rate of 0.34\,mag/hour for the rest 2.5 hours of observation \citep{gazeas14}. The MAGIC telescopes
started observations of the transient event at 21:08:54 UT, about 30\,s after the alert was received, and kept collecting data for the next 3.3\,h. A preliminary
analysis gave an integral flux upper limit of $1.2{\times}10^{-11}\,{\rm erg}\,{\rm cm}^{-2}\,{\rm s}^{-1}$ at ${\rm E}{\ge}200\,{\rm GeV}$ with a
confidence limit of 95\%, corresponding to 5.3\% of the Crab Nebula flux in this energy range, assuming a Crab-like spectral index \citep{mirzoyan14}.

Several more smaller flares were observed with UVOT and XRT \citep{drake14} after the initial trigger. They decreased in
peak brightness as the overall brightness decreased. When the {\it Swift} X-Ray
Telescope (XRT) started observing at ${\rm T}_{0}+117\,{\rm s}$, the soft X-ray 0.3-10\,keV rate of DG~CVn was ${\approx}100$\,${\rm cts}\,{\rm s}^{-1}$
and then decayed moderately, reaching a count rate of ${\approx}50$\,${\rm cts}\,{\rm s}^{-1}$ by ${\approx}328$\,s after the trigger.
The soft X-ray emission had declined to a level of 4-15\,${\rm cts}\,{\rm s}^{-1}$, but at ${\rm T}_{0}+11\,{\rm ks}$ {\dgcvn} was observed to have had a second, smaller
flare that peaked at ${\approx}30\,{\rm cts}\,{\rm s}^{-1}$ in the XRT detector. We will refer to this as the second flare hereafter.

\subsection{Observations with ``Pi of the Sky''}

\noindent The ``Pi of the Sky'' (PI) experiment is designed to monitor a large fraction of the sky with a high time resolution (10\,s) and self-triggering
capabilities \citep{cwiok07,mankiewicz14}. This means that PI may be performing observations of the field of sources well before the trigger time (${\rm T}_{0}$), the latter given by high-energy instruments (like {\it Swift}/BAT). This
approach resulted in the optical monitoring of {\dgcvn} even before ${\rm T}_{0}$. PI observed {\dgcvn} from ${\rm T}_{0}-700$\,s until ${\rm T}_{0}+1\,100$\,s. In
Fig.~1 the light curve is shown since ${\approx}{\rm T}_{0}-150$\,s onwards (since no significant brightness variations were detected before).

As can be seen, the optical flash only lasted about 60\,s, with the source brightening by over 4\,mag, to ${\rm V}{\approx}7$ at maximum
(occurring at ${\rm t}={\rm T}_{0}-41.3{\pm}0.4$\,s), and ending before the maximum of the hard X-ray emission which triggered the BAT alert (${\rm T}_{0}$). Then, a slow
decrease of the optical brightness was observed reaching ${\rm V}=11$ after about 0.5\,h.

\subsection{Observations with {\it BOOTES}}

\noindent {\it BOOTES} (acronym of the {\it Burst Observer and Optical Transient Exploring System}) is a world-wide network of robotic telescopes. It
was originally designed from a Spanish-Czech collaboration that started in 1998 \citep{castro99,castro12}. The telescopes are located
in Spain ({\it BOOTES}--1, {\it BOOTES}--2 and {\it BOOTES}--IR), New Zealand ({\it BOOTES}--3) and China ({\it BOOTES}--4).
Currently, one optical spectrograph is built and working in the {\it BOOTES} network, i.e. {\it COLORES} at {\it BOOTES}--2. {\it COLORES} stands
for {\it Compact Low Resolution Spectrograph} \citep{rabaza14}. It works in the wavelength range of
($3\,800-11\,500$)\,${\rm \AA}$ and has a spectral resolution of ($15-60$)\,${\rm \AA}$. The primary scientific target of the spectrograph is prompt GRB
follow-up, particularly for the estimation of redshift, but it is also used to study optical transients.

{\it COLORES} is a multi-mode instrument that can switch from imaging a field to spectroscopy by rotating wheel-mounted grisms, slits and filters
within an otherwise fixed optical system. {\it BOOTES}--2/{\it COLORES} started observing {\dgcvn} at ${\approx}{\rm
T}_{0}+11$\,min and
started to take spectra later at ${\approx}0.04$\,d (i.e. ${\approx}1$\,h)
after ${\rm T}_{0}$, following the evolution of the source for several months. In this
paper we report only on the observations when the source showed emission Balmer lines (i.e. when it was ``active''). See
Tab.~1 and 2 and Fig.~3 for a log of the observations and the
evolution of the spectra during the period of activity from the source (${\rm t}-{\rm T}_{0}{\le}5$\,d).

{\it BOOTES}--4 observed {\dgcvn} from ${\rm t}-{\rm T}_{0}=450$\,s to ${\rm t}-{\rm T}_{0}{\approx}0.01$\,d continuously in the clear and the Sloan g',r',i' and
the UKIDSS Z,Y filters. The
best resolution achieved was 0.5\,s. {\dgcvn} showed a monotonic decrease in flux during this period. Magnitudes close to the quiescent value were obtained from
${\rm t}-{\rm T}_{0}{\approx}0.7$\,d onwards.

\begin{figure*}
\includegraphics[bb=14 14 466 360,width=16cm,angle=0,clip]{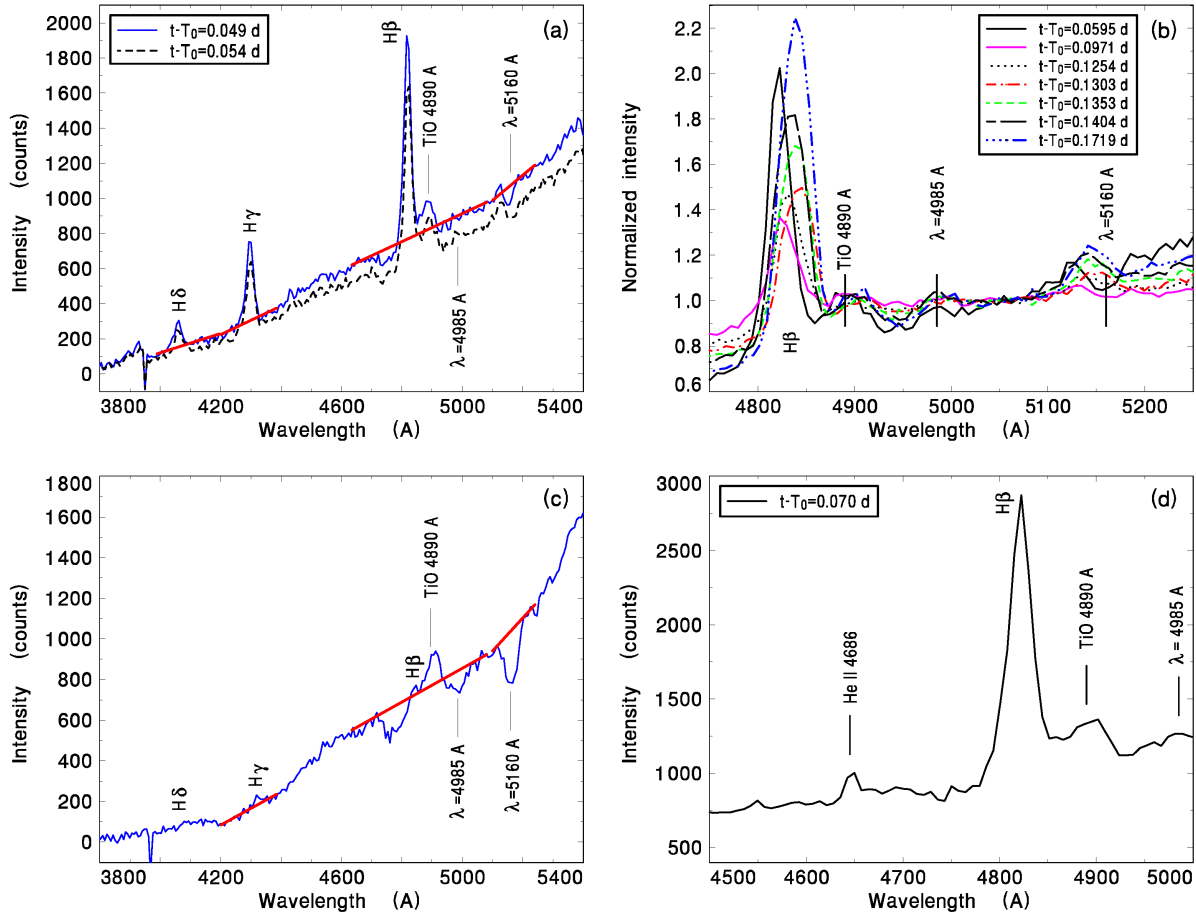}
\caption{{\bf {\it BOOTES}--2/COLORES spectra taken before, during and after the second flare event}. {\bf a}, The spectra were taken
at ${\rm t}-{\rm T}_{0}=0.049,0.054$\,d. These show intense Balmer emission
lines (${\rm H}_{\beta}$, ${\rm H}_{\gamma}$ and ${\rm H}_{\delta}$) and some absorption-emission features at ${\lambda}=4\,890,4\,985,5\,160\,{\rm \AA}$. The regions
used for the fitting of the pseudocontinuum are marked with the red line; {\bf b}, Spectra obtained during the time of the flare. Time-evolution of the (relative to the continuum) intensities
of the ${\rm H}_{\beta}$ Balmer and transient emission line at ${\lambda}{\approx}5\,160\,{\rm \AA}$;
{\bf c}, Spectrum at the late phase of the burst (${\rm t}-{\rm T}_{0}=2.976$\,d). Emission Balmer
 lines are marginally detected and the strength of the absorption features (${\lambda}=4\,890,4\,985,5\,160\,{\rm \AA}$) has increased. The feature at
${\lambda}=5160\,{\rm \AA}$ underwent a large change in time -- from a blue-wing emission and a shallow absorption (panel a) to a gradual increase of absorption which even replaced the emission line (panel c). This line was in a deep absorption in the
following nights. The intensity is normalized to unity at ${\lambda}=5\,050\,{\rm \AA}$ in panel c.{\bf d}, Spectrum showing the He~II ($4\,686\,{\rm \AA}$) emission line taken close 
to the time of the second flare (${\rm t}-{\rm T}_{0}=0.07$\,d).
}
\end{figure*}

\subsection{Observations with GTC/OSIRIS and BTA/SCORPIO}

\noindent We observed {\dgcvn} with the {\it Optical System for Imaging and low-intermediate Resolution Integrated Spectrograph} (OSIRIS; \citealt{cepa00}), located at the focal 
plane of GTC~(10.4\,m), on
2014--04--23 (see Tab.~1). We used three grisms with three different spectral resolutions,
i.e. R\,1000\,R, R\,2500\,R and R\,2000\,B. The blue spectra, R\,2000\,B, cover the (3\,950-5\,700)\,${\rm \AA}$ energy range and intense Balmer emission lines can be seen
(${\rm H}_{\beta}$,${\rm H}_{\gamma}$,${\rm H}_{\delta}$ and ${\rm H}_{\epsilon}$) at (4\,861,4\,341,4\,102,3\,970)\,${\rm \AA}$, respectively.

In the {\it OSIRIS} red spectra, the R\,1000\,R and R\,2500\,R grisms cover the energy ranges (5\,100-10\,000),(5\,575-7\,685)\,${\rm \AA}$, respectively.
They are very similar to the spectra obtained
with {\it BOOTES-2}/COLORES, with the most noticeable features being the presence of molecular absorption bands and the
intense ${\rm H}_{\alpha}$ emission line, at ${\approx}6\,563\,{\rm \AA}$, typical for the spectra emitted by ``active'' late-spectral type stars.

We also obtained a medium-high spectral resolution spectrum with the {\it Spectral Camera with Optical Reducer for Photometric and Interferometric Observations} 
(SCORPIO; \citealt{afanasiev05}), mounted
on the BTA (6\,m) telescope from the {\it Special Astrophysical Observatory} (SAO) on May 8$^{\rm th}$, 2014. Less intense Balmer emission lines can be seen. This indicates that
the source still maintained a (low) level of activity at this time.

\subsection{Observations with CAHA/CAFE}

\noindent We observed {\dgcvn} with the {\it Calar Alto Fiber-fed Echelle Spectrograph} (CAFE), located at the focal plane of the Calar Alto (CAHA)~(2.2\,m) telescope, on
2014--06--15 (see Tab.~2). This is a high (${\rm R}{\approx}62\,000{\pm}5\,000$) resolution spectrograph working in the $(3\,960-9\,500)\,{\rm \AA}$) wavelength range.
The spectral resolution is $<0.01\,{\rm \AA}$.

The high-resolution spectrum of {\dgcvn} during 2014--06--15, at the 2.2\,m CAHA ($+$CAFE), contains plenty of absorption lines (mainly TiO bands), with few
emission lines. The only instances of significantly broad (absorption) lines are cited in the following. Balmer lines are not seen either
in absorption or in emission. Indeed, the spectrum is typical of a cold late-type (``non-active'' at that time) star. There are prominent
broad absorption lines at $7\,970-8\,010\,{\rm \AA}$, that dominate the spectrum in this energy range. They correspond to VO molecular transitions. These
molecular transitions appear only for late-type M dwarf stars (M7 or later; \citealt{keenan52,kirkpatrick91}). Therefore the red-dwarf might be smaller than previously 
considered. Nevertheless, a proper spectral classification of the emitting star on the basis of the observed high-resolution spectral features is out of the scope of this
paper.

We observe at $8\,662\,{\rm \AA}$ the Fe I doublet in absorption. The presence of this doublet is indicative of a late spectral type star in a ``non-active'' state at that time.
This absorption doublet line disappears when the star turns activity on, due to the chromospheric heating produced during the stellar flares \citep{martin99}.
The activity of {\dgcvn} is greatly diminished during our CAFE observations and this translates into deeper Fe I and absorption lines in general.

\begin{table}
 \centering
 \begin{minipage}{120mm}  \caption{Log of the photometric observations reported in this work.}
  \label{log_1obs}
  \begin{tabular}{@{}lcc@{}}
  \hline
   Obs. Date$^{1}$  &     ${\Delta}\,{\rm t}^{2}$    &  Telescope/Instrument              \\
\hline
2014-04-23 -- 2014-04-23        &  -0.002 -- 0.003              &  PI                          \\
2014-04-23 -- 2014-04-27        &  0.0075 -- 3.305              &  BOOTES-2                    \\
2014-04-23 -- 2014-04-24        &  0.005 -- 0.759               &  BOOTES-4                    \\
\hline
\end{tabular}
\footnotetext { $^1$ The start time of an observation.}
\footnotetext { $^2$ The elapsed time from the start of the GRB (in days).}
\end{minipage}
\end{table}

\begin{table}
 \centering
 \begin{minipage}{120mm}  \caption{Log of the spectral observations reported in this work.}
  \label{log_2obs}
  \begin{tabular}{@{}lcc@{}}
  \hline
   Obs. Date$^{1}$  &     ${\Delta}\,{\rm t}^{2}$    &  Telescope/Instrument              \\
\hline
2014-04-23 -- 2014-04-27        &  0.044 -- 3.305               &  BOOTES-2/COLORES            \\
2014-04-25                      &  1.232 -- 1.236               &  GTC/OSIRIS-R1000R           \\
2014-04-25                      &  1.233 -- 1.238               &  GTC/OSIRIS-R2500R           \\
2014-04-25                      &  1.238 -- 1.246               &  GTC/OSIRIS-R2000B           \\
2014-05-08                      &  14.87 -- 14.88               &  BTA/SCORPIO-I               \\
2014-06-15                      &  52.99                        &  CAHA/CAFE                   \\
\hline
\end{tabular}
\footnotetext { $^1$ The start time of an observation.}
\footnotetext { $^2$ The elapsed time from the start of the GRB (in days).}
\end{minipage}
\end{table}

\section{Results}  \label{results}

\subsection{Optical and X-ray emission during the first flare}  

\noindent The delay of the initial peak of the X-ray emission ({\it Swift}/BAT in the 15-25\,keV energy range) with respect to the peak in the optical is clearly seen in
Fig.~1. We ascribe these observations to the so-called Neupert effect \citep{neupert68}, observed before (apart from the Sun) in UV~Ceti and
Proxima Centauri \citep{guedel96,guedel02,guedel04} with radio and X-ray observations of normal stellar flares. From the Sun and these stars it was seen that the X-ray light curve is observed to follow
approximately the time integral of the V-band emission (radio emission in the case of the Sun). The interpretation
is given in the framework of the {\it chromospheric evaporation} \citep{neupert68}, i.e. high-energy electrons travel along magnetic
fields, where the high-pitch angle population emits prompt gyrosynchrotron emission and the low-pitch angle population impacts in the chromosphere to
produce prompt radio/V band emission. The hot thermal plasma (soft X-rays) evolves as a consequence of the accumulated energy deposition, hence
the integral relation. The novelty in our case is that this relationship is observed to happen in the hard X-ray emission as well, contrary to previous expectations.

In the {\it chromospheric evaporation} the soft X-ray emission is a signature of the thermal emission from the heated plasma. This plasma is heated after the impact by
the accelerated particles, that are responsible for the early optical/radio-emission. Therefore, the detection of hard X-ray emission following the integral of the impulsive optical
emission is something unexpected. This indicates that, contrary to what has been previously understood, either the plasma heats-up to ${\rm E}{\ge}15$\,keV or that
the particles emit radiation following a non-thermal kinetic distribution.

\subsection{Optical spectroscopic observations}  

\noindent In the following we report on the optical spectroscopic observations performed with {\it BOOTES}--2/{\it COLORES} from the beginning of the second flare (approx.) onwards.
The optical spectrum from {\dgcvn} is characterized by a strong and variable red continuum with broad molecular TiO, CaI, MgI, NaI absorption lines/bands superimposed.
There are also intense Balmer emission lines. We notice that the intensities of the lines reported in this paper are not absolute fluxes, but
relative values with respect to the continuum. To calculate the relative intensities reported in this work, local continuum segments were taken at
${\lambda}=3\,990-4\,195,4\,200-4\,385,4\,635-5\,080,5\,100-5\,240\,{\rm \AA}$ for each ${\rm H}_{\beta}$, ${\rm H}_{\gamma}$ and ${\rm H}_{\delta}$
emission Balmer line, respectively. We label the continuum calculated from these segments ``pseudocontinuum'',
since it is not the true continuum, because it is very much affected by absorption band features in the spectrum (not constant a priori).
We considered these Balmer emission lines
to be excellent indicators of the spectral variability compared to the ${\rm H}_{\alpha}$ line. We did this in order to avoid possible effects due to
(minimal) saturation, because the ${\rm H}_{\alpha}$ line was very bright. The ``relative'' intensity of these lines (with respect to the continuum) is variable with time, showing
a constant ``plateau'' during the early phase of the burst (${\rm t}-{\rm T}_{0}{\le}0.10$\,d), with the exception of a ``depression'' (i.e. broad minimum at
${\rm t}-{\rm T}_{0}{\approx}11$\,ks) that coincides both in time and duration with the second flare event (Fig.~4). After the ``plateau'', the
(relative) intensity of the Balmer lines progressively decreased with time until ${\rm t}-{\rm T}_{0}{\approx}3$\,d. Eventually they remained minimally oscillating
around a mean value close to unity (i.e. the continuum flux level).

The long term evolution of the spectrum can be roughly characterized by the decrease of the intensity of the Balmer emission lines. At the early stage of the burst
(${\rm t}-{\rm T}_{0}{\le}3$\,d) these lines are prominent and clearly visible in the spectrum. These lines show that the star is chromospherically ``active''
at the moment they are present. Eventually (${\rm t}-{\rm T}_{0}>15$\,d), these lines disappear and the spectrum becomes emission-featureless. This is the
typical spectrum of a ``non-active'' cold red-dwarf star.

But apart from the intense Balmer emission lines and broad molecular bands there are further weaker absorption and emission lines. We based our identification
of these lines on the similarities with those shown in the spectra from cold red-dwarf stars \citep{pettersen85}. We find a number of spectral line features, at
${\lambda}=4\,890,4\,985,5\,160\,{\rm \AA}$. The first and the second are emission-like and the third shows an emission$+$absorption-line feature. After looking in more
detail into the time evolution of the ${\lambda}=4\,890\,{\rm \AA}$ emission-like line we realized that this feature is close to the boundary of the TiO absorption band. The
feature at ${\lambda}=4\,985\,{\rm \AA}$ looks like a weak emission that is only sometimes present. The ${\lambda}=5\,160\,{\rm \AA}$ feature is often seen in absorption but sometimes
it looks like a combination of absorption line and emission on its blue wing. These two lines correspond to TiO ($4\,985\,{\rm \AA}$) and MgH+TiO ($5\,160\,{\rm \AA}$) absorption bands, when compared with
the higher resolution OSIRIS R2000B spectra taken during the 3rd day of {\it BOOTES}--2 observations. Nevertheless, still marginal residual emission
at $5\,160\,{\rm \AA}$ is seen in the OSIRIS R2000B (we will discuss this feature hereafter). Also time evolution of the CaOH depression (at
${\lambda}=5\,530\,{\rm \AA}$) is shown in Fig.~4. The presence of the CaOH depression is a way to distinguish normal M stars from Me (emission)
stars \citep{pettersen85}. The CaOH depression was weaker in the early phase of the burst (${\rm t}-{\rm T}_{0}{\le}0.1$\,d) and then it increased in the late phase
(${\rm t}-{\rm T}_{0}>0.2$\,d). A wider broad dip minimum in the strength of the CaOH similar to that seen for Balmer lines at ${\rm t}-{\rm T}_{0}{\approx}0.12$\,d is
also observed for CaOH.

In Fig.~4 we show that both ${\rm H}_{\beta}$ line and the CaOH absorption band decrease in relative strength at the time of the second flare. The time
evolution of the Balmer lines and the CaOH absorption line are correlated during the early
phase of the burst ${\rm t}-{\rm T}_{0}{\le}0.2$\,d. Close to the end of the burst, the Balmer emission lines become marginal and the CaOH absorption band maximal. We will discuss
hereafter that both CaOH and Balmer lines features are good tracers of the behaviour of the
continuum from the source. In Fig.~4, we show that the line at ${\lambda}=4\,890\,{\rm \AA}$ (TiO) is the least time variable feature. We have checked whether this feature
could be a good indicator for the local continuum taken around ${\rm H}_{\beta}$ and found similar values to those obtained when using the local pseudo-continuum. This makes
us confident on our choice of the local continua when calculating line (relative) intensities.

There are also transient emission features in our spectra at ${\lambda}{\approx}5\,160,5\,850,4\,647\,{\rm \AA}$. These lines are compatible with blue-shifted Mg~I, He~D3 and He~II
(at ${\lambda}=5\,183,5\,876,4\,686\,{\rm \AA}$, respectively). They are blue-shifted by ${\approx}15\,{\rm \AA}$ (${\approx}37\,{\rm \AA}$ in the case of He~II), i.e. the same 
amount of shift as the ${\rm H}_{\alpha}$ line
(as shown hereafter), thus indicate chromospheric origin like the Balmer emission lines. Indeed, these lines have been seen during solar total eclipses \citep{voulgaris10}, supporting their
chromospheric origin. Also, the $5\,876\,{\rm \AA}$ line appears in emission in very strong solar flares (\citealt{johns97} and references therein). Apart from this, these transient 
lines are
absent during most of our spectral observations. Namely, these lines show an increase of their intensity
during ${\rm t}-{\rm T}_{0}{\ge}10$\,ks, coinciding with the time of the second flare event (see Fig.~3). During the rest of the time
the $5\,160\,{\rm \AA}$ feature becomes purely a MgH+TiO absorption feature. We will discuss on the importance of these transient
emission features hereafter.

\begin{figure}
\includegraphics[bb=0 0 468 360,width=7cm,angle=0,clip]{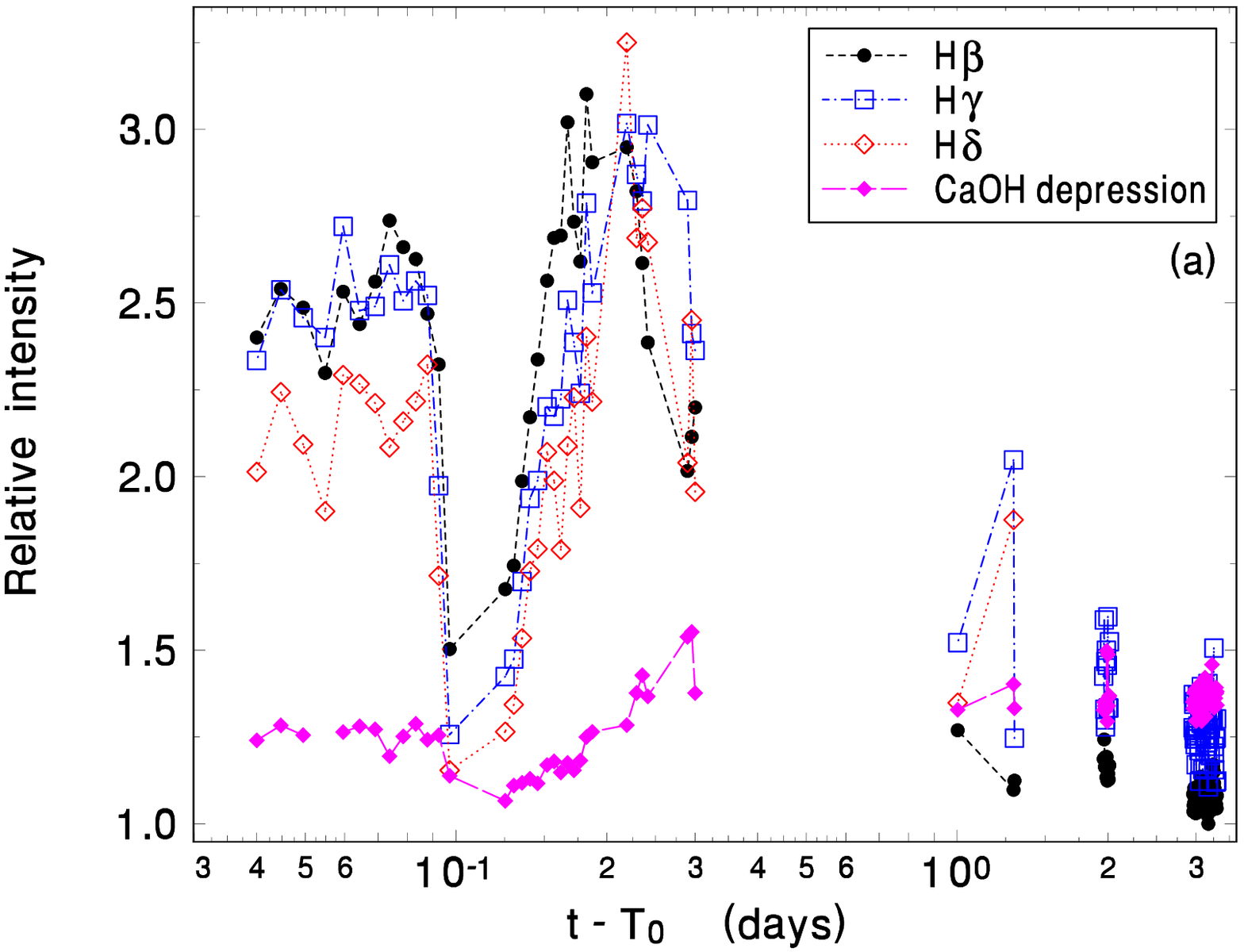}
\includegraphics[bb=0 0 468 360,width=7cm,angle=0,clip]{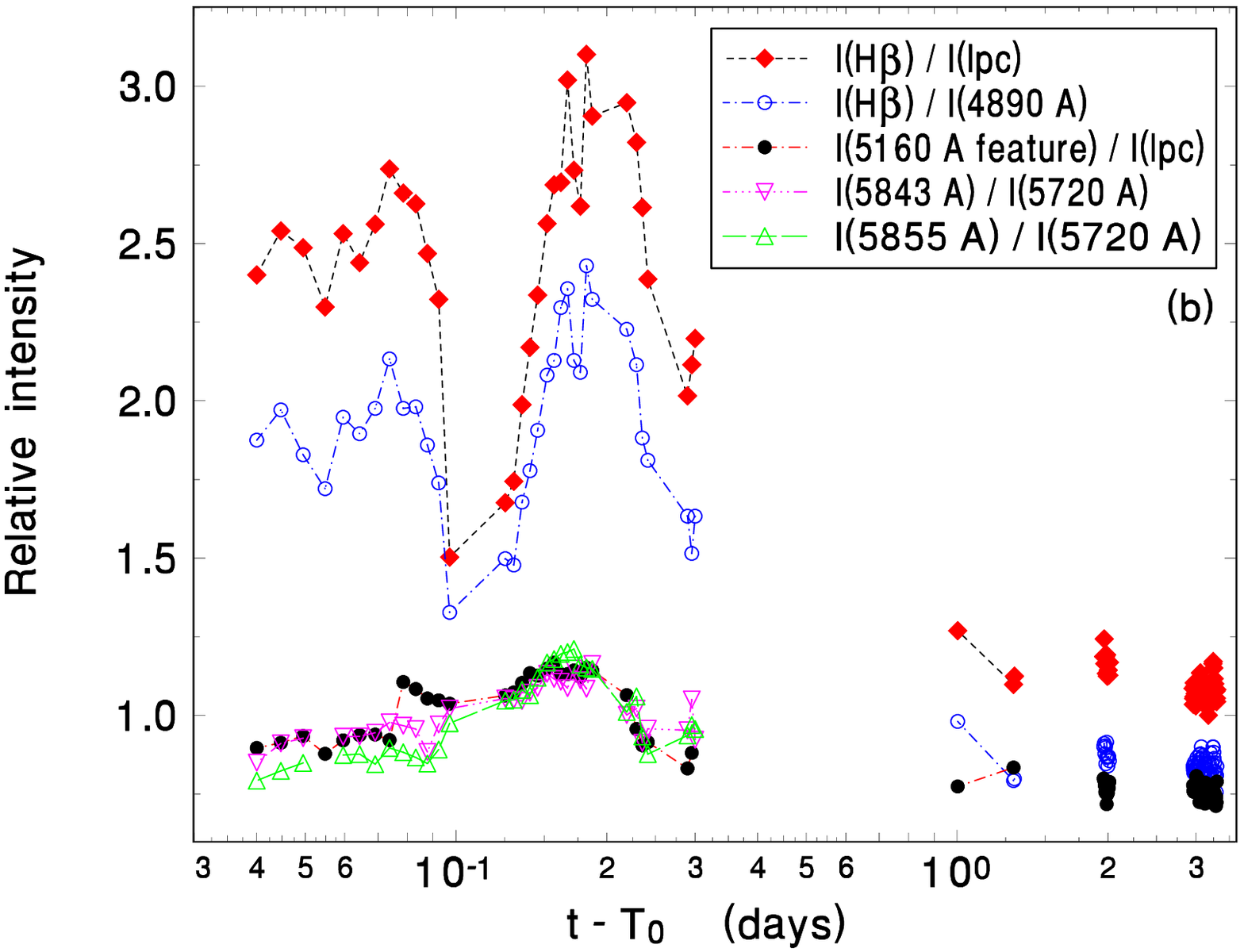}
\includegraphics[bb=0 0 468 360,width=7cm,angle=0,clip]{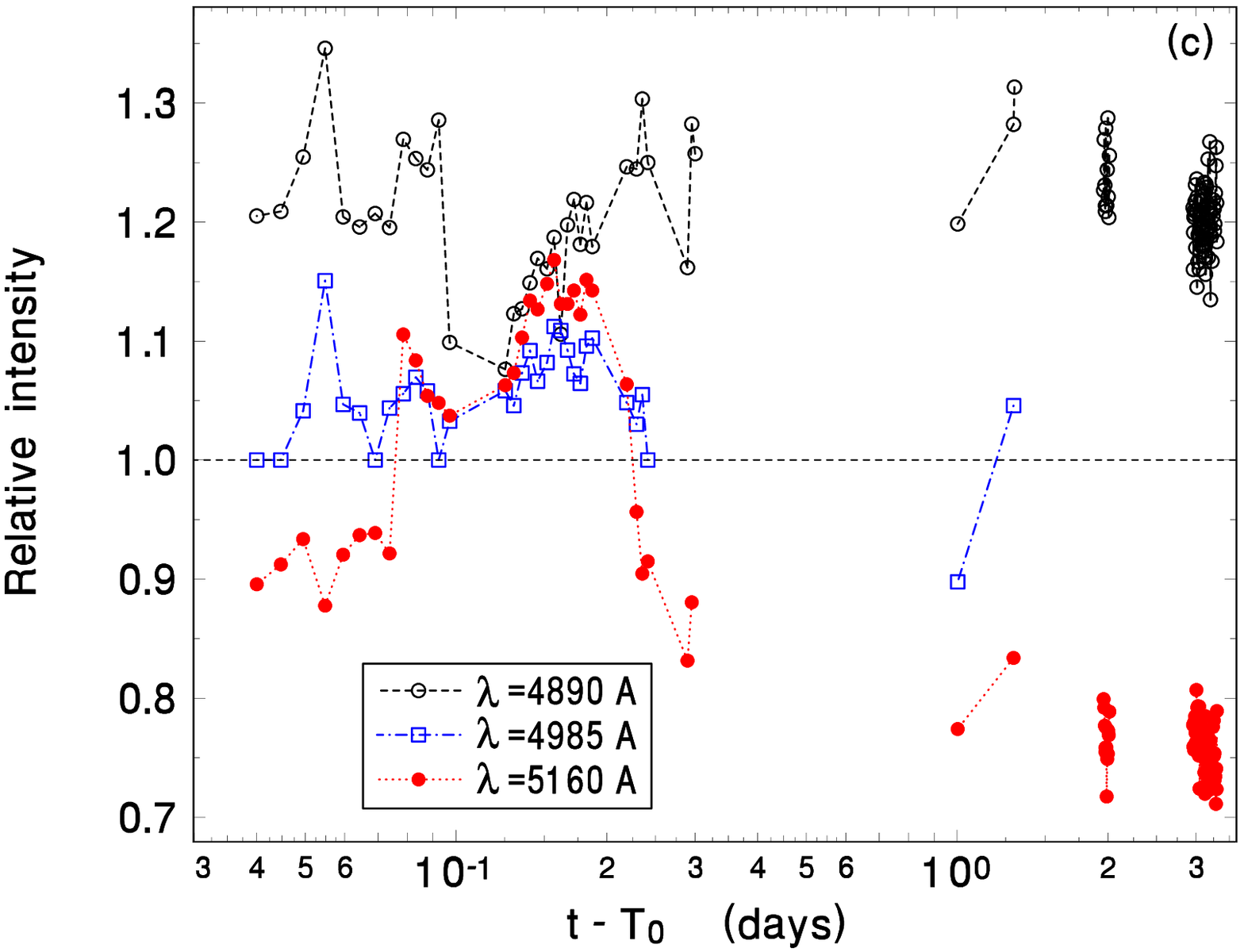}
\caption{{\bf Evolution of the (relative to the continuum) line intensities of the {\it BOOTES}--2/COLORES spectra, fully covering the second flare event}. {\bf a}, peak intensities of the Balmer lines measured with respect to the local pseudocontinuum. These Balmer lines appear in strong emission
shortly after ${\rm T}_{0}$ and this emission decreases in the next days. There is a minimum of Balmer line emission at ${\rm t}-{\rm T}_{0}{\approx}0.12$\,d, coinciding with
the time of the second smaller flare event. CaOH depression was the deepest near ${\rm t}-{\rm T}_{0}=0.13$\,d and it was the shallowest near ${\rm t}-{\rm T}_{0}=0.3$\,d;
{\bf b}, the time evolution of the peak intensities
of ${\rm H}_{\beta}$ and the ${\lambda}=5\,160\,{\rm \AA}$ emission feature are similar although the transient
decrease of the relative intensity at ${\rm t}-{\rm T}_{0}{\approx}0.12$\,d is considerably
smaller for the ${\lambda}=5\,160\,{\rm \AA}$ feature. The time evolution of the relative intensity of ${\rm H}_{\beta}$ is mutually similar no
matter if it is related to the local pseudocontinuum or to the ${\lambda}=4\,890\,{\rm \AA}$ feature (justifying that the local continuum is a good measure of the relative intensity
of the line); {\bf c}, time evolution of
the intensities of several spectral features (measured with respect to the local continuum). These features vary less than Balmer lines. The uncertainties of the
peak intensities are comparable to the size of the symbols.
}
\end{figure}

\subsection{Shifts of the ${\rm H}_{\alpha}$ line}  

\noindent Looking into the {\it COLORES} red spectra of {\dgcvn} (in a period out of X-ray activity) in more detail, we see that the emission ${\rm H}_{\alpha}$ line is shifting and we 
can tentatively see that it has a double-peaked profile. This is better seen in the higher resolution {\it OSIRIS} (Fig.~5) and {\it SCORPIO} spectra. The ${\rm H}_{\alpha}$ line shows
a complex P-Cyg profile, with two emission components
neither of which are centred at the nominal centre of an ${\rm H}_{\alpha}$
emission line. The brightest and the faintest peaks are red and blue-shifted by variable amounts, with the centre of the lines moving in the
(6\,572-6\,579)\,${\rm \AA}$ and (6\,548-6\,555)\,${\rm \AA}$ wavelength ranges, respectively. The centre of the complex (i.e. absorption component) of the two lines is
static ($6\,563{\pm}1$)${\rm \AA}$, coinciding with the centre of the ${\rm H}_{\alpha}$ emission line at rest.

The double profile ${\rm H}_{\alpha}$ has ${\approx}15{\rm \AA}$ red-shifted and a blue-shifted components. If we interpret these shifts as moving material
around {\dgcvn}, it would be moving in and out of the star at a speed of $500\,{\rm km}\,{\rm s}^{-1}$. This is far greater
than the previously measured rotational speed of the star (i.e. $50\,{\rm km}\,{\rm s}^{-1}$) and indicates the likely presence of important chromospheric 
activity. Additionally, the stationary absorption line corresponds to photospheric absorption from the surface of the active star. Unfortunately, neither 
${\rm H}_{\beta}$, ${\rm H}_{\gamma}$ nor ${\rm H}_{\delta}$ show the same effect, due to their lower intensities.

Previous studies observed a solar flare with high-resolution spectroscopy found that the ${\rm H}_{\alpha}$ emission line during the flare was
also double-peaked during flaring activity periods \citep{johns97}. As in our case the line was also asymmetric, with the red wing having more emission 
than the blue. This effect was interpreted as redshifted emission coming from chromospheric condensations and falling towards
the photosphere, or else as blueshifted absorption coming from optically thin material and rising towards the corona. They found broad emission ${\rm H}_{\alpha}$
line profiles (${\approx}100\,{\rm km}\,{\rm s}^{-1}$). In our case the line profiles are broader (${\approx}500\,{\rm km}\,{\rm s}^{-1}$), indicating the presence
of a more extreme/powerful flare event giving rise to them.

\begin{figure}
\centering
 \includegraphics[bb=0 0 612 792,width=7cm,angle=270,clip]{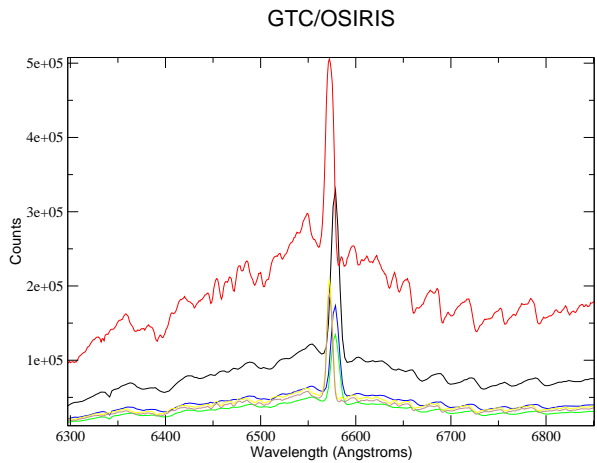}
 \caption{GTC/OSIRIS spectra from Obs.~1--6 measured from the data taken on the 2014/04/25 night (in black, red, green, blue, yellow and brown, respectively).
}
 \label{plot_shift}
\end{figure}

\section{Discussion}   \label{discuss}

\subsection{On the estimation of the age of the system}   

\noindent The age of {\dgcvn} still has not been well established. Originally it was estimated to be between the ages of the stellar 
associations ${\beta}-{\rm Pic}$ (12\,Myr) and AB Dor (125\,Myr), based on
its gravity measurement and deblended H-R diagram positions \citep{riedel14}. Nevertheless, it has been considered to be a young system (30\,Myr) 
because its membership to one of the solar neighborhood 30\,Myr associations gives the best match \citep{riedel14}. {\dgcvn} and the 
other two fast rotators in the solar neighbourhood belong to the so-called zone ``C'' of fast rotators which is a mixture 
of stars spanning a wide range of ages and/or spectral types \citep{mohanty03}. Indeed, there is no consensus yet on the physical mechanism 
that causes a fraction of stars to retain rapid rotation rates compared to other stars of similar age (see \citealt{brown14} for a discussion on
possible physical scenarios).

To give a further insight into the age of {\dgcvn} we have computed non-stop evolution models from the PMS (pre-main sequence)
to WD (white-dwarf) cooling sequences for two masses: 0.2 and 0.3\,${\rm M}_{\odot}$. These masses are typical for red dwarfs with spectral types between 
M4V and M3V \citep{kaltenegger09}. The calculations were performed using the MESA code \citep{paxton11,claret12}. We adopted
an initial chemical composition around the solar value (X=0.70, Z=0.02) and a mixing-length parameter $\alpha_{MLT}$ = 1.89. The commonly accepted age of 30\,Myr is neither 
compatible with main
sequence nor near zero-age main sequence (ZAMS) models since the corresponding radii are too large and the star would be still in the PMS phase. However,
if we assume that DG~CVn contains a star
which is at the ZAMS or near it, we find that the radius agrees, within uncertainties, with those tabulated \citep{kaltenegger09}. The inferred age in this case is about
150\,Myr for typical M4V spectral type (${\ge}150$\,Myr if the star is more evolved than zero-age main sequence). The same conclusions are inferred for a star of
spectral type M7V or later. We notice that the previously reported age ($30$\,Myr; \citealt{riedel14}) is not wrong and agrees with our estimation only if we consider that
it is the age of the system after achieving the ZAMS (which usually takes a time of ${\approx}100$\,Myr of gravitational contraction from the original stellar cloud).

\subsection{Conclusions}   

\noindent In this paper we have studied the spectral evolution of {\dgcvn} during an episode of important optical and hard X-ray activity. During this period at least 
two flares have been detected
to occur quasi-simultaneously in both energy bands. We have explained the origin of the (few tens of seconds) hard X-ray delay with respect to the optical emission (Fig. 1) by the Neupert
effect, which was observed before (apart from the Sun) in UV Ceti and Proxima Centauri during normal
soft X-ray flares. The novelty of our study is that we observe this effect, but for the hard X-ray emission. The X-ray luminosities during the peaks of emission are
${\rm L}_{\rm X}(0.3-10)\,{\rm keV}=1.4{\times}10^{33},3.1{\times}10^{32}\,{\rm erg}\,{\rm s}^{-1}$. Although the flare X-ray luminosities are certainly high
similar values (${\rm L}_{\rm X}(0.3-10)\,{\rm keV}{\ge}10^{32}\,{\rm erg}\,{\rm s}^{-1}$) have been detected for a dozen of cases 
in main-sequence red-dwarf stars. Recently these values have been superseded by an intense episode of activity from the 
red-dwarf binary system SZ~Psc \citep{drake15}. Also the duration of the flaring episode is no exception, with similar values reported for the
(9\,day) flare of CF~Tuc observed by \citet{kurster96} and the previous super-flare from EV~Lac \citep{osten10}. Such episodes of increased X-ray/optical activity
have never been observed in the Sun so far. However, the same physics underlying the giant X-ray/optical solar flares
serves to explain the superflares observed in active-like stars (as {\dgcvn}; \citealt{aulanier13}). Nevertheless, since the most powerful solar flares have been of
only about $10^{32}\,{\rm erg}\,{\rm s}^{-1}$ \citep{carrington859} the scale-up of solar flares models would require enormous starspots (up to 48 degrees in latitude/longitude extent)
to match stellar superflares, thus much bigger than any sunspots in the last 4 centuries of solar observations. Therefore \citet{aulanier13} conjecture that one condition for
Sun-like stars to produce superflares is to host a dynamo that is much stronger than that of the Sun. It is also worth to mention that some recent studies 
\citep{kitze14,candelaresi14,wu15} point out to both the (possible) few instances and the possibility of Solar-type stars undergoing superflares with luminosities as high as 
$10^{35}-10^{37}\,{\rm erg}\,{\rm s}^{-1}$ under certain conditions, by using data from the {\it Kepler} satellite \citep{koch10}.  

In our work optical spectral 
observations began one hour after the onset of the burst, thus allowing a detailed study of the spectral properties of
{\dgcvn} during the evolution of the second flare. Balmer emission lines are prominent during the whole period of activity of the star. The optical spectra are similar to
the spectra emitted by ``active'' late-spectral type stars. A significant decrease in the (relative to the continuum) intensity of the Balmer lines is observed during the second flare (Fig.~2).
Balmer lines represent the formation and evolution of the optically thin medium above the stellar photosphere (i.e. the chromosphere). On the other hand, the variable
strengths of the CaOH and TiO features represent the variations of the photosphere of {\dgcvn} itself (where the absorption lines and the continuum emission are thought
to come from). The Balmer and CaOH depressions are due to an increase of the continuum emission as well. The presence of absorption bands is
due to the molecules present in the photosphere of {\dgcvn}. As the continuum level increases the (relative to the continuum) strength of the Balmer
lines and the CaOH absorption band decreases, as observed during the second flare. Therefore, the
decrease of intensity of these spectral features is not real, but an artifact of an increase of the surrounding continuum.

It is in the chromosphere where the Balmer and the rest of emission lines originate. The chromosphere is more tenuous and hotter than the photosphere.
Therefore only emission lines from light elements can be seen, except in the coronal transition region, where the high temperatures can ionize Fe. An increase
in the intensity of the Mg~I\,($5\,183\,{\rm \AA}$), He~D3\,($5\,876\,{\rm \AA}$) (and He~II; $4\,686\,{\rm \AA}$) emission lines have been detected during the 
second flare of {\dgcvn}. This
indicates a correlation between the photospheric and the chromospheric activity. In other words, a change in the properties of the continuum is related to properties of the
chromosphere of the red-dwarf star. We will discuss in the following on the most likely origin of this change.

The super-flare observed from {\dgcvn} during April 2014 may correspond to an episodic high-intensity flare event from a red-dwarf star. This flare event is originated from one
or more active regions on the star. If there is only one active region giving rise to the observed flaring activity this star would be spinning very rapidly, with a period of a 
few hours (${\rm P}_{\rm rot}{\approx}2.5$\,h), thus giving rise to the optical phenomenology observed. Photometric studies of {\dgcvn} performed earlier have revealed a periodicity 
of ${\rm P}_{\rm phot}=0.10836(2)$\,d \citep{robb99}. Indeed, \citet{robb99} claimed that a hot spot model fit their single-band data best. This period is
very close to the elapsed time between the first and the second flares{, which might argue in favour of one hot spot being the responsible for both flares}. Nevertheless, the 
presence of a second active region (responsible for the second flare) can not be discarded. Indeed, many active stars have active 
regions separated by 180 degrees in longitude. This would ``double'' the rotational period of the star. Also, given that rapidly rotating stars often have large
polar spots, the lack of a dimming of the second flare due to rotational modulation might also be due to the fact that the associated active region(s) responsible for the
optical flare might be also circumpolar.  

Further spectroscopic and photometric observations of {\dgcvn} would help in order to identify which is (are) the active component(s), their inclination and the actual 
rotation of the star. The use of spatially resolved measurements and multi-band photometric data would help us to understand the exact origin of the variability.

\section*{Acknowledgments}

To the memory of J. Gorosabel, for his large dedication and contribution to the study of cosmic gamma-ray emitting sources of unknown origin. We thank the anonymous referee 
for helpful comments. This work was supported by the European social fund within the framework of realizing the 
project ``Support of inter-sectoral mobility and quality enhancement of research teams at Czech Technical University in Prague'', CZ.1.07/2.3.00/30.0034. MCG thanks
to B. Montesinos, H. Krimm, T. Sakamoto and S. Pedraz for discussions. AJCT, MJ and SRO thank the
support of the Spanish Ministry Projects AYA2009-14000-C03-01 and AYA2012-39727-C03-01. ACw, RO and AFZ acknowledge financial support
received from the Polish Ministry of Science and Higher Education. RH acknowledges GA CR grant 13-33324S. Analysis of the data was
partially based on the software developed within the GLORIA project, funded from the European Union Seventh
Framework Programme (FP7/2007-2013) under grant agreement n$^{\circ}$ 283783.

\end{document}